\documentclass[letterpaper, 10 pt, conference]{ieeeconf}
\IEEEoverridecommandlockouts
\title{\LARGE \bf
Analysis of the Rising Threat of Subverting Privacy Technologies
}

\author{Craig Ellis}

\begin{document}

\maketitle
\thispagestyle{empty}
\pagestyle{empty}

\begin{abstract}

Privacy technologies have become extremely prevalent in recent years from secure communication channels to the Tor network. These technologies were designed to provide privacy and security for users, but these ideals have also led to increased criminal use of the technologies. Privacy and anonymity are always sought after by criminals, making these technologies the perfect vehicles for committing crimes on the Internet. This paper will analyze the rising threat of subverting privacy technologies for criminal or nefarious use. It will look at recent research in this area and ultimately reach conclusions on the seriousness of this issue.

\end{abstract}

\section{INTRODUCTION}

In recent years, privacy technologies have flourished and have received more attention. With the growth of the Internet, more and more users have sought out the ability to privately and securely communicate. These technologies are designed to provide anonymity and privacy for their users including whistleblowers and others who need to be able to communicate anonymously. Along with these new technologies came a rise in cybercrime. Anonymity is something that cybercriminals always desire and these technologies provide an easy way to achieve it. This led to an explosion in untraceable ransomware, Tor botnets, and other crimes that cannot easily be tracked back to an attacker. With the rise of Bitcoin, payments can also be made anonymously and cannot be tracked back to the attacker. The privacy infrastructure was designed to help users, but also led to this steep rise in cybercrime and a steep increase in the difficulty of catching these cyber criminals. This paper looks at how this privacy infrastructure is subverted for attacks and how serious of a threat this poses to the Internet as a whole. Section II will outline a few privacy technologies, Section III will outline the research analyzed within this paper, and Section IV will present conclusions drawn from this research.

\section{PRIVACY TECHNOLOGIES}
As usage of the Internet has exploded, the need for security and privacy when using the Internet has also increased dramatically. To meet these rising needs, many new privacy technologies have been created from secure protocols, like TLS/SSL, to Tor, a network that allows for anonymous Internet usage, to Bitcoin which allows anonymous payments to be sent over the Internet. These technologies were all created to help provide security and privacy over the Internet in order to allow for services to be provided online. Whistleblowers and individuals living under totalitarian regimes can share information, individuals can anonymously browse the Internet, and individuals can check their bank accounts and make purchases online. The ideals that these technologies aimed to provide were positive, but were also very useful for cyber criminals, leading to many crimes and attacks that subvert these privacy technologies for nefarious means.

\subsection{SSL/TLS}
SSL/TLS is a technology that was designed to allow for secure communication across the Internet. It is a protocol that was built on top of TCP that involves a session between a client and a server and then connections related with that session. It allows for one or two way authentication, confidentiality, and message integrity, all of which make the communication more secure and less vulnerable to attacks. SSL/TLS has been widely deployed in order to provide these features to users, especially with HTTPS, which allows for secure browsing of the Internet and the ability to securely make payments over the Internet.

\subsection{Tor}
Tor is a network that aims to provide anonymous Internet browsing and communication. It is built on the idea of Onion routing, which involves messages passing through multiple servers in such a way that each node only knows the information about the previous and next nodes. In this infrastructure, the first node will only know about the user and second node, not the destination, that the last node will only know about the destination and the previous node, not the sender, and that intermediate nodes only know about the previous and next nodes. This is done by encrypting messages with a key for each node so that as the messages maneuver through the network, each layer of encryption is peeled off. Tor also allows decoupling of the IP address and the address used to visit the website so that the server hosting the site cannot easily be tracked down. 

Tor does not provide end to end security, as it only encrypts the information within the network, so if end to end security is needed, it is provided by the end server. There have been a few recorded vulnerabilities in the Tor network, but it is widely considered to be very secure. The foundation of the Tor network led to the ability to browse the Internet and communicate anonymously, with the latter being very important to government and military entities, but, along with this, came an increase in sophisticated cybercrime, like ransomware and secure botnets, and the ability to create websites that offer illicit drugs, murder for hire, and many other illegal activities.

\subsection{Bitcoin}
Bitcoin is the first successful cryptocurrency. It is a form of currency that is completely electronic and is not backed by some other means like a government. By design, it allows for Bitcoins to be transferred anonymously but also securely. It is based on a chain of transactions, which is used to ensure that nobody is lying about a transaction and allows for Bitcoins to be transferred. Bitcoins are also only tied to an address, called a wallet, which is not tied directly to an individual's identity. Since the currency is not connected to an identity, a transaction cannot be tracked back to see who it came from. The ability to make anonymous payments over the Internet can be very useful, but can also be used in many crimes. One major crime that became popular with the rise of Bitcoin is ransomware, which is malware that will infect a device, encrypt important files, and then require payment in order to get the key that will unlock the files. With the invent of Bitcoin, attackers can require that payment is made in the form of Bitcoins, making the payment untraceable. Bitcoin also lead to an increase in other crimes, like purchasing illegal drugs and even hiring hitmen, since the payment cannot be traced.

\section{RESEARCH ON SUBVERTING PRIVACY TECHNOLOGIES}

\subsection{Botnet over Tor: The Illusion of Hiding [1]}

This research was performed by Casenove and Miraglia from VrijeUniversiteit in Amsterdam, The Netherlands. It analyzes how botnets try to guarantee anonymity of the Command and Control node, addresses the problems with Tor-based botnets, and shows how full anonymity is never achieved.

The researchers found that as botnets began to make use of the Tor network, they typically used it to hide the connection, as best as possible, from infected devices to the command and control node. This was successful in its intended purpose of hiding the communication, but the act of using the Tor network often led to abnormal traffic that revealed the fact that a device was infected with a botnet. This was ultimately counterproductive to the ultimate purpose of the botnet; while the control servers were more difficult to find and take down, it was more obvious that the infected devices were infected, which makes them easier to fix. The researchers then discuss the fact that botnets that make use of Tor are still susceptible to many attacks that affect non-Tor based botnets. Crawling the address space to determine how many hosts are infected is still feasible and so is using traffic analysis at the exit nodes to determine the address of a command and control node. Using Tor in this manner has made it more difficult to track down the command server, but it does not solve all issues relating to botnets, but as these botnets evolve and make better use of Tor, they will be increasing difficult to undermine.

This research demonstrates that botnets that make use of privacy technologies, like Tor, are already here and are becoming more and more sophisticated. The bots analyzed in this research were only making minimal use of Tor and were not attempting to hide the fact that Tor was running on the infected devices, yet they were still successful when only using privacy technologies minimally. As these botnets become more sophisticated, they will become more difficult to track and take down. They will be better able to hide their communications and hide themselves on the infected devices. While botnets that make use of Tor are relatively new, they will continue to improve and use more privacy features, highlighting the threat that comes from subverting privacy technologies and the need to get ahead of these types of attacks.

\subsection{OnionBots: Subverting Privacy Infrastructure for Cyber Attacks [2]}

This research was performed by Sanatinia and Noubir from Northeastern University in Boston, MA. The goal of this research was to introduce a method by which a botnet could operate using the Tor infrastructure and then propose a method for neutralizing bots that are using the Tor network.

The proposed method for creating a botnet that uses the Tor network relies on a self-healing, low degree, and low diameter graph of bots. New nodes would join the botnet by contacting a list of current bots and then generating a connection between itself and that bot. This would be done in such a way that the bots would have a small number of neighbors, but the shortest path to any other bot would also be small. These nodes would also maintain the network by healing when a node leaves the graph. Since these nodes are using the Tor network, there is a decoupling between the IP address of a node and its .onion address. This allows the nodes to continually change their .onion address while still allowing neighbors to be able to communicate to it and allowing the Command and Control node to control it. The research shows that this botnet would be able to maintain a low degree and diameter while also not becoming partitioned if many nodes drop out of the network. It would also be difficult to track and take down since it is built on the Tor network and the devices can frequently change their .onion addresses in order to make finding the infected devices more difficult.

The research then proposed a method for defeating these so-called OnionBots. The technique is called Sybil Onion Attack Protocol - SOAP. The basis of the technique is that a node controlled by a defender that is part of the botnet is able to get a neighboring node to drop all of its other links and connect only to nodes the defender set up. The defender does this by creating new nodes that are under its control (replicas) that then connect to the target node. Part of the design of the botnet states that a node should connect to the neighboring nodes with the lowest degree and drop its neighbors with the highest degree in order to do so. The defender can exploit this by having the new nodes connect to the target node and report having very few neighbors. This will cause the target node to drop its neighbors and connect to the defender controlled nodes. Once the targeted node is only connected to the defender owned nodes, it is then disconnected from the botnet and is neutralized.

The goal of this research was to show that it is possible to create a botnet that operates using the Tor network for privacy and anonymity and show that there are ways to counteract this type of botnet. The threat posed by botnets operating over the Tor network is severe. Botnets have already been used in many crimes and attacks across the Internet, including the DDoS attack launched in October 2016 that targeted Dyn and prevented access to numerous websites. Botnets are already difficult to track down and stop, but botnets that make use of the Tor network would be extremely difficult to take down and could degrade service within the Tor network as a whole, leading to issues for its intended users. As the previous research showed, Tor based botnets already exist, and while they are not very sophisticated yet, this research shows that this class of botnets can continue to improve and become more successful. Ultimately, the research argues that these issues should be dealt with preemptively, including making changes to these privacy technologies if needed. Subverting these privacy technologies is a significant threat to any user of the Internet and a preemptive strategy is the best strategy that can be adopted at this point as it would protect against these technologies being used for crimes and can still provide the services they provide to users who are using them for non-criminal means. If the threat of subverting Tor for botnet use is not dealt with preemptively and quickly, it could become an enormous issue for the entire Internet and lead to a significant increase in untrackable crimes.

\subsection{Preparing for Malware that Uses Covert Communication Channels: The Case of Tor-based Android Malware [3]}

This research was performed by Kioupakis and Serrelis from AMC Metropolitan College in Amaroussio, Greece. It proposes a methodology for creating Android based malware that uses the Tor network to extract and steal information from the device and then proposes a method to mitigate this type of malware. 

The researchers first discuss the fact that there are already two known cases of Android based malware that made use of the Tor network in some capacity, although it was typically used to retrieve information from the Command and Control bot. After analyzing these previous botnets, they proposed a design, including requirements and module implementations, for an Android based malware that makes use of the Tor network to steal information from the device. This malware targets the most popular version of Andoird, Jelly Bean, and would not be caught by anti-malware implementations that currently exist. It would be able to sit on Android devices, undetected, and siphon information that is then transmitted back to a central server, possibly compromising personal and confidential information that is contained in or sent from the device.

After describing in detail how the malware is built, they propose an anti-malware system that can help to detect and stop malware that makes use of the Tor network. This anti-malware system would sniff packets off of the network and would shut down an interface if it is sending Tor traffic and it is not in a list of devices approved to use the Tor network. Current malware mitigations would not be able to detect the malware they designed, so new techniques were required that could detect and stop this type of malware. They propose that this type of anti-malware be developed preemptively in order to detect and stop future malware variants that will make use of Tor.

The research demonstrates that Tor can be used for many purposes within the broad scope of cybercrime and the security community needs to begin to take note and get ahead of these uses. If malware is able to run on Android devices undetected and anonymously send data back to a Command and Control server, then this is a very large threat to an enormous number of users. Android has one of the largest user bases in the world, so any malware that affects these devices could be extremely effective and lucrative. A reactive approach to this issue could lead to the loss of personal information for millions of users before this malware is identified and stopped. Tor has many useful purposes, but the protections it provides to cyber criminals must be minimized in order to protect Tor and the Internet's future use, even if it means making modifications to the Tor network or the devices that use it. Malware that uses privacy technologies is already here and will continue to become more stealthy and secure, demonstrating the need to attack the issue now before it becomes even more prevalent.

\subsection{Ransomware: Emergence of the Cyber-Extortion Menace [4]}

This research was performed by Hampton and Baig from the Security Research Institute at Edith Cowan University in Perth, Australia. The goal of this research was to show the trends that have occurred in the history of ransomware and make obvious the idea that security professionals need to get ahead of these issues instead of constantly being behind and reacting to the issue.

The researchers analyzed twenty-nine variants of ransomware that belonged to nine different families. For each variant, they looked for twenty-two different features that these variants could possess to see what features have been kept and also see what recent trends are emerging. They found that cryptographically secure encryption was not used until around 2013 and in recent years, there has been an increase in the use of Tor and cryptocurrencies like Bitcoin. The researchers suggest that the limited data makes it difficult to make predictions, but history has shown that future generations learn from the success and failure of past generations, and as of now, using Tor and other privacy technologies seems to be successful. Their analysis suggests that future variants of ransomware will make more use of privacy technologies in order to protect the cybercriminals and make it that much more difficult to track them down.

The researchers suggest, similar to the research described above, that security researchers need to get ahead of the trends instead of reacting as they occur. This is not possible in every situation, but if researchers can look towards the future and try to predict what might occur, it can be mitigated before it becomes a larger issue. They also describe the fact that vulnerability analysis has become a systematic and well vetted process with numerous databases detailing almost every imaginable vulnerability, yet malware, specifically ransomware, analysis has not had the same prominence and success. Ransomware is currently a massive business bringing in millions of dollars per year for cybercriminals and it affects individuals, businesses, and the government alike. The researchers ultimately advocate for creating a formal approach for documenting and analyzing malware, especially ransomware, so that security professionals can prevent more malware variants from springing up, which has been happening rapidly in the past few years. Once this process is in place, security researchers can begin to counteract these strains of malware, including reducing the ability of these strains to use privacy technologies for nefarious means.

\section{Conclusions}

As all of this research has shown, the possibility of subverting privacy technologies for nefarious means is very real. These attacks are not just a future possibility; they are here now and appear to be extremely successful. These technologies, like Tor and Bitcoin, were designed to provide users with a sense of privacy and anonymity when using the Internet. Tor was designed to help people communicate in an anonymous and untraceable way, which is great for military use of for whistleblowers who are afraid of persecution. Bitcoin was developed to allow people to make monetary transactions that were not tied to their identity and could not be traced back to them. These technologies were all designed and implemented with the intent to help people, but they are also being used to commit crimes and even help terrorists.

Addressing these issues is a real concern that must be met with proactive measures. Far too many times, especially with security, we are reactive to situations. Many protocols and, in fact, the basis of the Internet was designed without security in mind and it was only added when bad things began to happen. There was a swift reaction to these issues and those problems were fixed. In order to stop the viral spread of cybercrime, made much easier with these technologies, there needs to be a proactive plan. As the research from Hampton and Baig stated, we need to create a system of standardization and understanding when it comes to malware and cybercrime research. We also need to continue to push for more research like the research presented here. This research looked at future possibilities and explored what may be coming in order to make suggestions of how it can be stopped now. If we, as a community, can look ahead and understand how issues may arise in the future, we can fix them now and stop the current rise of cybercrime, and also prevent new variants from popping up in the future. 

The privacy technologies discussed in this paper were designed for good reason and provide many useful services, but in order for them to be truly successful and usable, there need to be mitigations in place that prevent cybercriminals from exploiting the protections provided. The research above discusses future threats that undermine privacy technologies, like Tor and Bitcoin, but the threat of subverting privacy technologies for criminal means is here now and must be dealt with in a proactive manner in order to stop these crimes and prevent them from appearing in the future to the greatest extent possible.

\end{document}